\documentclass[aps,twocolumn,showpacs,preprintnumbers,prb,superscriptaddress]{revtex4-1}

\usepackage{graphicx}
\usepackage{amssymb}
\usepackage{amsmath}
\usepackage{hyperref}

\newcommand{\Eq}[1]{Eq.~(\ref{#1})}

\newcommand{\Ref}[1]{Ref. \cite{#1}}
\newcommand{\Fig}[1]{Fig. \ref{#1}}
\newcommand{\Tab}[1]{Tab. \ref{#1}}
\newcommand{\Sec}[1]{Sec. \ref{#1}}
\newcommand{\App}[1]{Appendix \ref{#1}}

\newcommand{\bra}[1]{\ensuremath{\langle #1 \vert}}
\newcommand{\ket}[1]{\ensuremath{\vert #1  \rangle}}

\usepackage{color}

\newcommand{\R}{\mathbf{r}}

\begin{document}
	
	\title{Methods to generate the reference total and Pauli kinetic potentials.}

	\author{Szymon \'Smiga}\email{szsmiga@fizyka.umk.pl}
\affiliation{Institute of Physics, Faculty of Physics, Astronomy and Informatics, Nicolaus Copernicus University, Grudziadzka 5, 87-100 Toru\'n, Poland}

	\author{Sylwia Sieci\'nska}
\affiliation{Institute of Physics, Faculty of Physics, Astronomy and Informatics, Nicolaus Copernicus University, Grudziadzka 5, 87-100 Toru\'n, Poland}

	\author{Eduardo Fabiano}
	\affiliation{Institute for Microelectronics and Microsystems (CNR-IMM), Via Monteroni, Campus Unisalento, 73100 Lecce, Italy}
	\affiliation{Center for Biomolecular Nanotechnologies @UNILE, Istituto 
		Italiano di Tecnologia (IIT), Via Barsanti, 73010 Arnesano (LE), Italy}

\begin{abstract}{We have derived a new method which allows to compute the full and the Pauli reference
kinetic potentials for atoms and molecules in a real space
representation. This is done by applying the optimized effective potential
(OEP) method to {the} Kohn-Sham non-interacting kinetic energy expression.
Additionally, we have also derived a simplified OEP variant based on the
common energy denominator approximation 
which has proven to give much more stable and robust results than the original
OEP one. Moreover, we have also proved that at the solution point our 
approach is formally equivalent to the commonly used Bartolotti-Acharya formula.}

\end{abstract}

\maketitle

\section{Introduction}\label{sec:intro}
The most natural and straightforward realization of density functional theory (DFT)
\cite{HK-teo,Levy6062} is the so called orbital-free (OF) DFT 
\cite{Wang2002,KARASIEV20122519}. The theory describes the ground-state electronic properties
of any electron system via the knowledge of the electron density $\rho$ that is
obtained as the solution of the Euler equation \cite{EulerEq} 
\begin{equation}\label{eq:euler}
\frac{\delta T_s[\rho]}{\delta \rho(\R)} + v_{ext}(\R) + v_H(\R) + \frac{\delta E_{xc}[\rho]}{\delta \rho(\R)} = \mu \; ,
\end{equation}
where $v_{ext}$ is the external/nuclear potential, $v_H$ is the Hartree potential, 
$\mu$ is a Lagrange multiplier fixed from the normalization condition 
$\int d\R \rho(\R) = N$, and $E_{xc}$ and $T_s$ are the exchange-correlation (XC) and 
non-interacting kinetic energy (KE) functionals, respectively. 
The latter two quantities are very important to describe many-body fermionic effects.
However, despite the existence of both functionals is guaranteed by 
the first Hohenberg-Kohn theorem \cite{HK-teo} and they can be formally defined within
the Levy's constrained search procedure \cite{Levy6062}, their explicit expression in
terms of the electron density is unknown. Therefore, one needs to approximate both 
quantities. 

In the case of the XC functional, many useful approximations have been proposed {(see e.g. 
\Ref{Li2018, MHGREV, della2016kinetic} and references therein)}. On the other hand, for the KE term, 
this task is much more difficult due to two facts: i) the KE contribution is 
much larger than the XC one (it has the same order of magnitude
as the total energy \cite{PhysRevA.32.2010,PhysRevA.26.1200}); ii) it includes highly 
non-local Pauli contributions 
\cite{dellasala15,Constatntin2016-KN,constantin2017jellium,constantin18},
 which account for all fermionic effects.
Nonetheless, different approximate KE functionals have been developed {
(see e.g. Refs. \cite{xia2015single,karasiev2006born,trickey2009conditions,karasiev09_chapter,karasiev2013nonempirical,luo2017differential,constantin2009kinetic,smiga2017laplacian,constantin2017jellium,smiga2015subsystem,constantin2017modified,seino2018semi,cancio2017visualisation,ernzerhof2000role,chakraborty2018kinetic,LC94,TW02,thakkar1992comparison,ou1991approximate,vitos2000local,lindmaa2014quantum,borgooJCTC13,yang86,lucianLL,fde_lap,CancioKin,RATIONAL1,Pavanello22} )}.
In numerous cases these have been developed mimicking the exchange functional construction,
according to the \emph{conjointness conjecture} hypothesis \cite{ConCon1,ConCon2,apbek, SmiComp2019}. 
In other cases, specific KE properties have been considered such as {the} exact constraint 
satisfaction \cite{karasiev2013nonempirical,luo18,trickey2009conditions} or the use of
information from the linear response of the uniform electron gas \cite{Wang2002}. 
In all cases, however, the approximate KE functionals show still in general poor accuracy and/or 
transferability \cite{seino2018semi,Carter2018,Luc-KIN}. 

For these reasons, OF-DFT calculations are rarely employed in practice. The
most popular computational realization of DFT, the Kohn-Sham (KS) method
\cite{kohn:1965:KS}, avoids the direct use of the KE functional by introducing an auxiliary 
non-interacting system of fermionic particles 
where both the density and the KE terms are expressed in terms of 
 single-particle orbitals $\{\phi_{i}\}$. Thus, we have
\begin{eqnarray}
\label{eq:kinrho}
\rho ({ \bf r}) &=& \sum_{i}|\phi_{i}({ \bf r})|^2 \\
\label{eq:kinrho1} 
T_s & =& -\frac{1}{2} \sum^N_i \int d\R \; \phi^*_{i}(\R)\ \nabla^2  \phi_{i} (\R) \ .
\end{eqnarray}
%
Note that the KE can be alternatively written as
\begin{equation}\label{eq4}
T_s =  \int d\R  \tau(\R)
\end{equation}
where $\tau (\R) = 1/2 \sum^N_i |\nabla \phi_{i}(\R) | ^2$  is the positive-definded KE density. 
{Throughout this paper we label with $i,j$ the occupied KS orbitals, with $a,b$ the unoccupied ones, with $p,q$ the general (occupied or unoccupied) ones. All equations are written in spin-restricted form.}

The KE functional given by  Eq. (\ref{eq:kinrho1}) is usually separated in two main contributions, 
namely the von Weizs\"{a}cker (VW) \cite{vW} ($T^W$) and {the} Pauli ($T^P$) term
\begin{equation}\label{eq:kinfunsep}
T_s[\rho] = T^W[\rho] + T^P[\rho] \; .
\end{equation}
The former has a simple semilocal expression written in terms of the density and 
its gradient, which reads
\begin{equation}\label{eq:kinfunsep1}
T^W[\rho]  =  \int d\R \frac{|\nabla \rho(\R)|^2}{8\rho(\R)} \; ,
\end{equation}
and is exact for any one- and two-electron spin-singlet state system{s}. 
Due to its explicit density dependence, the corresponding kinetic potential can be easily
derived using standard functional derivatives as 
\begin{equation}\label{eq:vWpot}
v^W (\R) = \frac{\delta T^W[\rho]}{\delta \rho(\R)} =   \frac{|\nabla \rho(\R)|^2}{8\rho^2(\R)}  - \frac{\nabla^2 \rho(\R)}{4\rho(\R)} \; .
\end{equation}
%

On the other hand, the Pauli term can only be expressed exactly via {the} KS orbitals \cite{TAl-IJQC-1978,Bar-JCP-1982,Levy-PRA-1988,StarINV} as  
\begin{equation}\label{eq:kinfunsep3}
T^P_s[\rho]  = \int d\R  ~\tau^P (\R) \; 
\end{equation}
%
with
\begin{equation}\label{eq:kinfunsep4}
\tau^P (\R) =  \frac{1}{2 \rho(\R)} \sum_{i<j} |\phi_{i}(\R) \nabla \phi_{j}(\R) - \phi_{j}(\R) \nabla \phi_{i}(\R) |^2 
\end{equation}
{beeing the Pauli kinetic-energy density according to \Ref{Levy-PRA-1988}.}

%

The KS method, via Eq. (\ref{eq:kinrho1}) [or Eq. (\ref{eq4})], provides a direct way to calculate the non-interacting KE of any electron system. Therefore, this information can be, and indeed is, used to assess and improve approximate KE functionals. On the other hand, the KS method makes no use of the kinetic potential, $\delta T_s[\rho]/\delta \rho(\R)$, and provides no direct way to obtain it. For this reason the kinetic potential has longly been an overlooked quantity and almost no effort has been made to assess and optimize the approximate KE functionals against this quantity. However, the kinetic potential is the main ingredient in the Euler equation [\Eq{eq:euler}] and its importance is nowadays increasingly recognized \cite{Hyd1,Hyd2,Wesol-2013,subDFT,Ast-2016,Carter2018,Mi-2018,Luc-KIN}.
Thus, methods to generate the kinetic potential from reference KS input quantities, or more generally from any input set of orbitals, are of great importance to allow a direct knowledge of this fundamental quantity.

The simplest way to generate the total kinetic potential is to use directly \Eq{eq:euler}. If the ground-state density $\rho(\R)$ is already determined and we fix the corresponding XC potential (e.g. from standard KS calculations \cite{kohn:1965:KS}, \textit{ab initio} DFT \cite{grabowski:2002:OEP2,bartlett:2005:abinit2,grabowski:2007:ccpt2,verma:044105,grabowski13,grabowski:2014:jcp,SCSIP,UHFOEP} or some ''reverse-engineering'' approach \cite{ZMP,wu:2003:wy,StarINV}), then 
{it is} clear that at the solution point the kinetic potential can be obtained as the negative of {the} effective potential shifted by a constant
\begin{equation}\label{eq:vkin1}
v_{\text{k}}(\R) = \frac{\delta T_s[\rho]}{\delta \rho(\R)}  = - v_{\text{s}}[\rho](\mathbf{r}) +  \mu \; ,
\end{equation}
where 
{with} $v_{\text{k}}(\R)$ we denote the total kinetic potential. The
constant $\mu$ is commonly taken \cite{Levy-PRA-1984,Alm-PRV-1985} to be the
negative of the first ionization potential $\mu = - IP$ or approximately the
orbital energy of the {highest occupied molecular orbital (HOMO)} ($\mu = \varepsilon_H$) which can be calculated 
{in} various manners\cite{SmigaIP2018}.

In order to compute the Pauli potential one simply needs to subtract the Weizs\"{a}cker kinetic potential given by \Eq{eq:vWpot} from \Eq{eq:vkin1} getting
\begin{equation}\label{eq:vkin1p}
v^P (\R)  = - v_{\text{s}}[\rho](\mathbf{r}) - v^W (\R) +  \mu \; .
\end{equation}
%

Another commonly used method to generate reference Pauli potentials \cite{Bar-JCP-1982,Levy-PRA-1988,FIn-IJQC-2016,Fin-JCP-2016,FIN-IJQC-2017,Const-PRB-2019} is the one derived by Bartolotti and Acharya (BA) in \Ref{Bar-JCP-1982}. The formula 
\begin{equation}\label{eq:BA1}
v^P(\R)  = \frac{\tau(\R)  - \tau^W(\R) }{\rho(\R) } + \sum^N_i \left(\varepsilon_H - \varepsilon_i \right)   \frac{|\phi_i(\R) |^2}{\rho(\R) }  \; 
\end{equation}
utilizes the occupied orbitals and {the} orbital energies from an arbitrary {self-consisted field (SCF)} method (including Hartree-Fock). The full derivation of \Eq{eq:BA1} can be found in \Ref{Bar-JCP-1982,Levy-PRA-1988}. However, for clarity of this paper it is also briefly sketched in \App{ap:ABF}.
 
Formally, \Eq{eq:BA1} is equivalent to \Eq{eq:vkin1p} with a corrected VW term (see \Ref{karasiev09_chapter}, \Sec{secvw} and \App{ap:ABF} for more details). 
Once the Pauli potential is available the total kinetic potential can be calculated as 
\begin{equation}\label{eq:BA2}
v_{\text{k}}(\R) = v^W(\R) +  v^P(\R)  \; .
\end{equation}
%

{
In this study,
we introduce a new method based on the optimized effective potential (OEP) \cite{sharp:1953:OEP,talman:1976:OEP} approach allowing to generate the non-interacting kinetic potentials
 (full and Pauli term{s})
for different atoms and molecules in a real space representation. Additionally, we compare and discuss the proposed method with {the} aforementioned approaches utilized up to date.}

\section{Theory}\label{sec:theor}
In this section
we introduce a new method based on the OEP technique that allows to generate the full and the Pauli kinetic potentials from any set of reference orbitals ($\phi_p$) and orbital energies ($\varepsilon_p$); in addition, we describe the common energy denominator approximation of the method that is numerically simpler and more stable.


\subsection{Kinetic potential using the OEP method}
\label{sec:kinoep3}
In this subsection we consider a new method to obtain the kinetic potential.
For simplicity we will describe it for the total kinetic functional; however, it can be
applied in exactly the same way also for the Pauli kinetic term 
(see \Eq{eq:kinfunsep4}).

To start consider the KS non-interacting kinetic energy functional given in 
\Eq{eq:kinrho1}. Because it displays an explicit orbital dependence, 
while it is only and implicit functional of density, 
the direct computation of {the} kinetic potential through functional derivative 
is impossible. Thus, in order to calculate the potential, alike in the case of 
the orbital dependent exchange 
\cite{sharp:1953:OEP,talman:1976:OEP,ivanov:1999:OEP,gorling:1999:OEP,KumelKronik,gorlingOEP2,ivanov:2002:OEP,EngelOEPx} 
and correlation 
\cite{grabowski:2002:OEP2,bartlett:2005:abinit2,grabowski:2007:ccpt2,verma:044105,grabowski13,grabowski:2014:jcp,SCSIP,EngelOEP2,EngelOEP3} 
energy functionals, one can employ the OEP method
\cite{sharp:1953:OEP,talman:1976:OEP}.
Hence, we can define the functional derivative of \Eq{eq:kinrho1} 
using following chain rule
 \begin{widetext}
\begin{eqnarray}\label{eq:oep2}
v_{k}({\bf r}) =   \sum_p
\int d{\bf r'}d{\bf r''} \left \{ \frac{\delta T_{s}[\{\phi_{q}\}]}
{\delta \phi_{p}({\bf r'})}
\frac{\delta \phi_{p}({\bf r'})}{\delta v_{s}({\bf r''})}
\frac{\delta v_{s}({\bf r''})}{\delta \rho({\bf r})}
  + c.c. \right \} \; .
\end{eqnarray}
 \end{widetext}
 
%
In the above equation the first term in brackets is easily derived (see \App{ap:GLK})
to be
{zero when $p$ indexes an unoccupied orbital, while for $p$ indexing an
  occupied orbital we find}
%

\begin{eqnarray}\label{eq:oep3}
\frac{\delta T_{s}[\{\phi_{q}\}]} {\delta \phi_{p}({\bf r'})} = -\frac{1}{2}  \nabla^2  \phi_{p} (\R') -  \frac{1}{2}  \nabla^2  \phi^*_{p} (\R')\ .
\end{eqnarray}
In case of real orbitals, as it often happens, 
this is just $-\nabla^2  \phi_{p} (\R')$.
The second term can be obtained from first-order perturbation theory considering an infinitesimal perturbation of {the} effective potential ($\delta v_s$) introduced into the KS equation.
Thus, we have
\begin{eqnarray}\label{eq:oep4}
  \frac{\delta  \phi_{p}({\bf r})}{\delta v_{s}({\bf r'})} =   \sum_{q \neq p} \frac{ \phi_{p}({\bf r'}) \phi^*_{q}({\bf r'})}{\varepsilon_{p}  - \varepsilon_{q}}  \phi_{q}({\bf r}) \; .
\end{eqnarray}
%
The last term, is the inverse ($X^{-1}(\R',\R)$) of the static KS linear response 
function of a system of non-interacting particles expressed trough 
KS orbitals and eigenvalues:
\begin{equation}\label{eq:oep4a}
X(\R',\R) = 2 \sum_{ia}\frac{\phi^*_{i}(\R')\phi_{a}(\R')\phi^*_{a}(\R)\phi_{i}(\R)}{\varepsilon_{i}-\varepsilon_{a}}\   + c.c. \; .
\end{equation}
%

Inserting \Eq{eq:oep3} and \Eq{eq:oep4} into \Eq{eq:oep2}, after some algebra we obtain
\begin{eqnarray}\label{eq:oep5}
&&v_{k}^{\mbox{\scriptsize OEP}}({\bf r}) =\\
&& \sum_{i,a} \Bigg[ \frac{ (T_s)_{i a}}{\varepsilon_{i}  - \varepsilon_{a}} \int d{\bf r'} \phi_{i}({\bf r'}) \phi^*_{a}({\bf r'}) X^{-1}({\bf r,r'})
  + c.c. \Bigg]  \nonumber  \; .
\end{eqnarray}
where $(T_s)_{p q} = \langle \phi_{p} \vert -\frac{1}{2} \nabla^2 \vert  \phi_{q} \rangle$ are the KE matrix elements. Note that {the} above procedure is partially similar to the one used in the self-consistent implementation of meta-GGA XC functionals depending on the local kinetic energy density \cite{della2016kinetic,metaGGA1,metaGGA3,metaGGA2}.

\subsection{Common energy denominator approximation}
\label{sec:kinoep4}
Since the OEP procedure described above is numerically involved and not very
stable (see \Sec{sec:res} for more details), we introduce here an approximation
based on the common energy denominator method {(CEDA)} \cite{CEDA,LHF,KLIapp}
This leads to a simpler and well behaving equation for the kinetic potential that
yields basically the same results as the full OEP variant (\Eq{eq:oep5}).

To obtain our approximation we start by multiplying \Eq{eq:oep5} by 
\Eq{eq:oep4a} and integrating over $\R$ to obtain
\begin{eqnarray}\label{eq:kli1}
&& \sum_{ia} \Bigg[ \frac{ (v_{k})_{i a}}{\varepsilon_{i}  - \varepsilon_{a}} \phi_{i}(\R) \phi^*_{a}(\R) 
  + c.c. \Bigg]  \\ 
&& =  \sum_{ia} \Bigg[ \frac{ (T_s)_{i a}}{\varepsilon_{i}  - \varepsilon_{a}} \phi_{i}(\R) \phi^*_{a}(\R) 
  + c.c. \Bigg] \nonumber  \; ,
\end{eqnarray}
%
%
%
where 
\begin{equation}\label{eq:kli2}
\left( v_{k} \right)_{p q}= \int d\R \phi_{p}^*(\R) v_{k}({\R}) \phi_{q}(\R)\ .
\end{equation}
This is just another representation of OEP equation\cite{OEPdiffform}.
Now we assume that all the energy differences in the denominator of \Eq{eq:kli1} 
can be approximated by a constant mean energy 
($\Delta \approx \varepsilon_{i}  - \varepsilon_{a}$) getting 
\begin{eqnarray}\label{eq:kli3}
&& \sum_{ia} \Bigg[ (v_{k})_{i a} \phi_{i}(\R) \phi^*_{a}(\R) 
  + c.c. \Bigg]  \\ 
&& =  \sum_{ia} \Bigg[ (T_s)_{i a} \phi_{i}(\R) \phi^*_{a}(\R) 
  + c.c. \Bigg] \nonumber  \; .
\end{eqnarray}
At this point we can use on both sides of \Eq{eq:kli3} the relation 
\begin{eqnarray}\label{eq:kli4}
\sum_{a} \phi^*_{a}(\R)  \phi_{a}(\R') + \sum_{i} \phi^*_{i}(\R)  \phi_{i}(\R') = \delta (\R - \R')
\end{eqnarray}
to obtain
\begin{eqnarray}\label{eq:kli5}
&& \rho (\R) v_{k}(\R)   - \sum_{i,j} (v_{k})_{i j} \phi_{i}(\R) \phi^*_{j}(\R) + c.c. =    \\
&& -\frac{1}{2} \sum_i \phi^*_{i}(\R) \nabla^2 \phi_{i}(\R)   - \sum_{i,j} (T_{s})_{i j} ~ \phi_{i}(\R) \phi^*_{j}(\R) + c.c. \nonumber \; .
\end{eqnarray}
Finally, using the identity \cite{della2016kinetic} 
\begin{equation}\label{tauleq}
-\frac{1}{2}\sum_i^N\phi_{i}^*(\R)\nabla^2\phi_{i}(\R) = \tau(\R)-\frac{1}{4}\nabla^2\rho(\R)
\end{equation}
%
we find the kinetic potential approximation 
\begin{eqnarray}\label{eq:kli6}
&& v_{k}(\R)  = \frac{\tau(\R)}{\rho (\R)} - \frac{\nabla^2 \rho (\R)}{4 \rho (\R) }  \\ 
&& + \sum_{i,j} \left[ (v_{k})_{i j} - (T_{s})_{i j} \right]
  \frac{\phi_{i}(\R) \phi^*_{j}(\R)}{\rho (\R)} \ . \nonumber \; 
\end{eqnarray}
This equation expresses the kinetic potential in the {CEDA}.

A further approximation can be obtained following the idea of Krieger-Li-Iafrate (KLI) \cite{KLIapp}, neglecting
in the summation all the terms with $i \neq j$. 
{Indeed, numerical investigations support the fact that the off-diagonal terms in the sum provide only a minor contribution with respect to the diagonal ones.}
In this way we obtain the KLI approximation of \Eq{eq:oep5} which reads 
\begin{eqnarray}\label{eq:kli7}
&& v_{k}(\R)  = \frac{\tau(\R)}{\rho (\R)} - \frac{\nabla^2 \rho (\R)}{4 \rho (\R) }  \\ 
&& + \sum_{i} \left[ (v_{k})_{i i} - (T_{s})_{i i} \right] \frac{|\phi_{i}(\R)|^2}{\rho (\R)}  \nonumber \; .
\end{eqnarray} 
Inserting the definition 
$\tau({\bf r}) = \tau^W({\bf r}) + \tau^P({\bf r})$ into \Eq{eq:kli7}
and removing the VW potential of Eq. (\ref{eq:vWpot}) one obtain{s} the Pauli potential 
\begin{eqnarray}\label{eq:kli8}
&& v^P(\R)  = \frac{\tau^P(\R)}{\rho (\R)}  + \sum_{i} \left[ (v_{k})_{ii} - (T_{s})_{i i} \right] \frac{|\phi_{i}(\R)|^2}{\rho (\R)}\; .
\end{eqnarray} 
%
One can prove that 
{for density and orbitals corresponding to the SCF solution of the KS and
  Euler equations (i.e.} 
at the solution point {)} the above formula is formally equivalent to \Eq{eq:BA1} (see \App{ap:ABF3} for more details).

\begin{figure}
\includegraphics[width=\columnwidth]{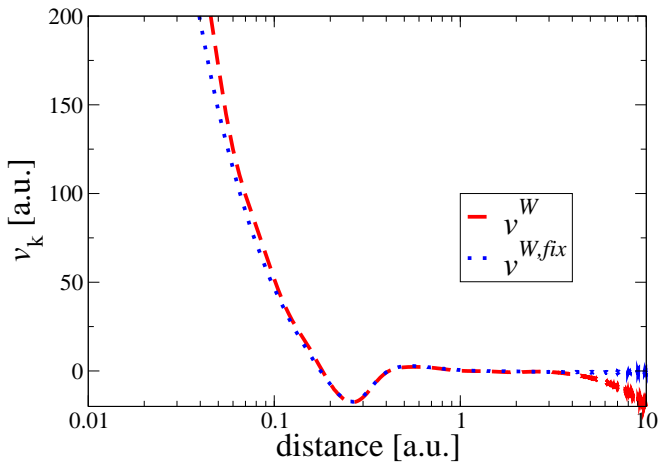}
\caption{\label{fig1} The Weizs\"{a}cker kinetic potential without and with oscillatory profile correction calculated for Ne atom in the cc-pVTZ basis set.}
\end{figure}

\begin{figure*}
	\includegraphics[width=1\columnwidth]{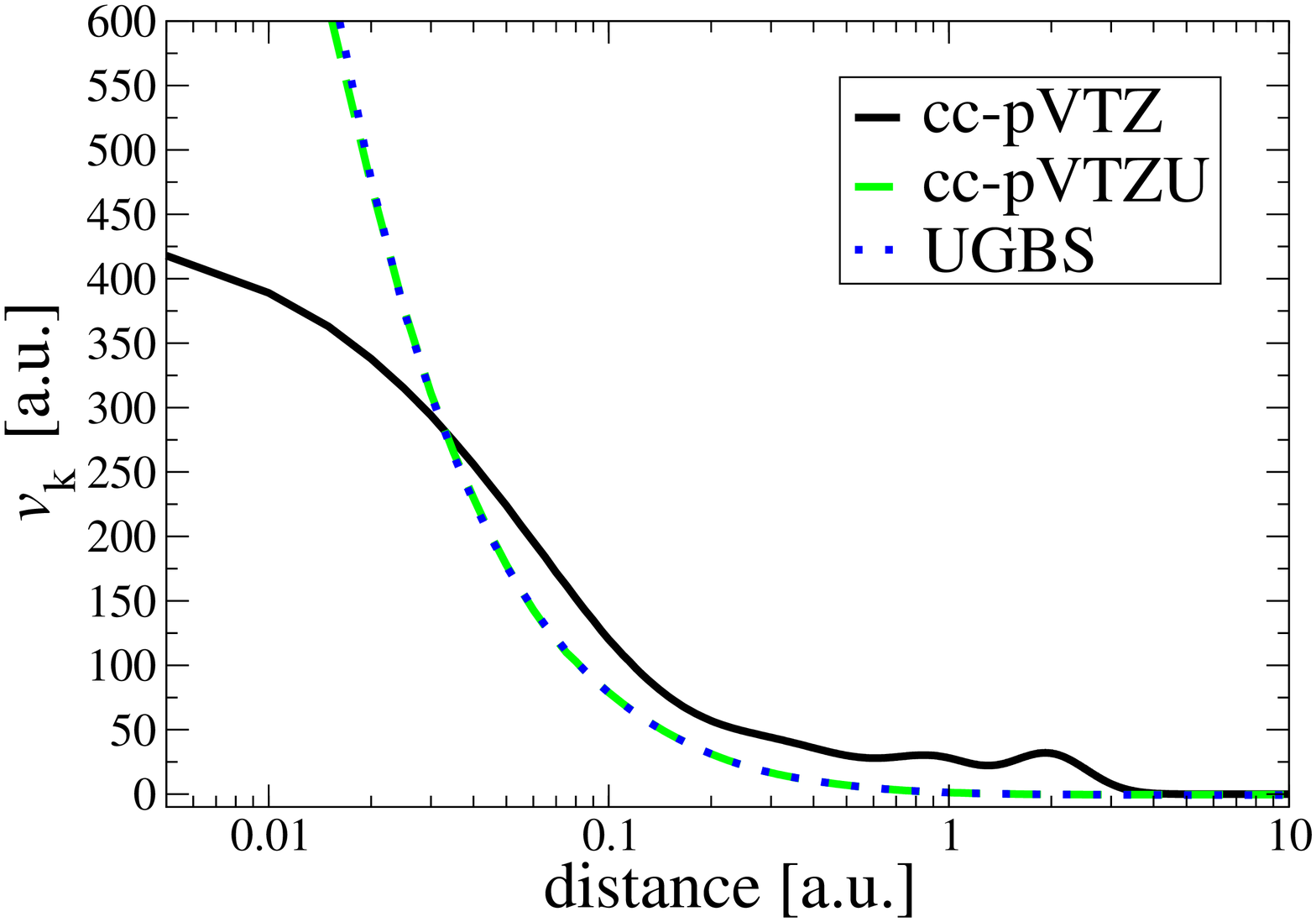}
	\includegraphics[width=1\columnwidth]{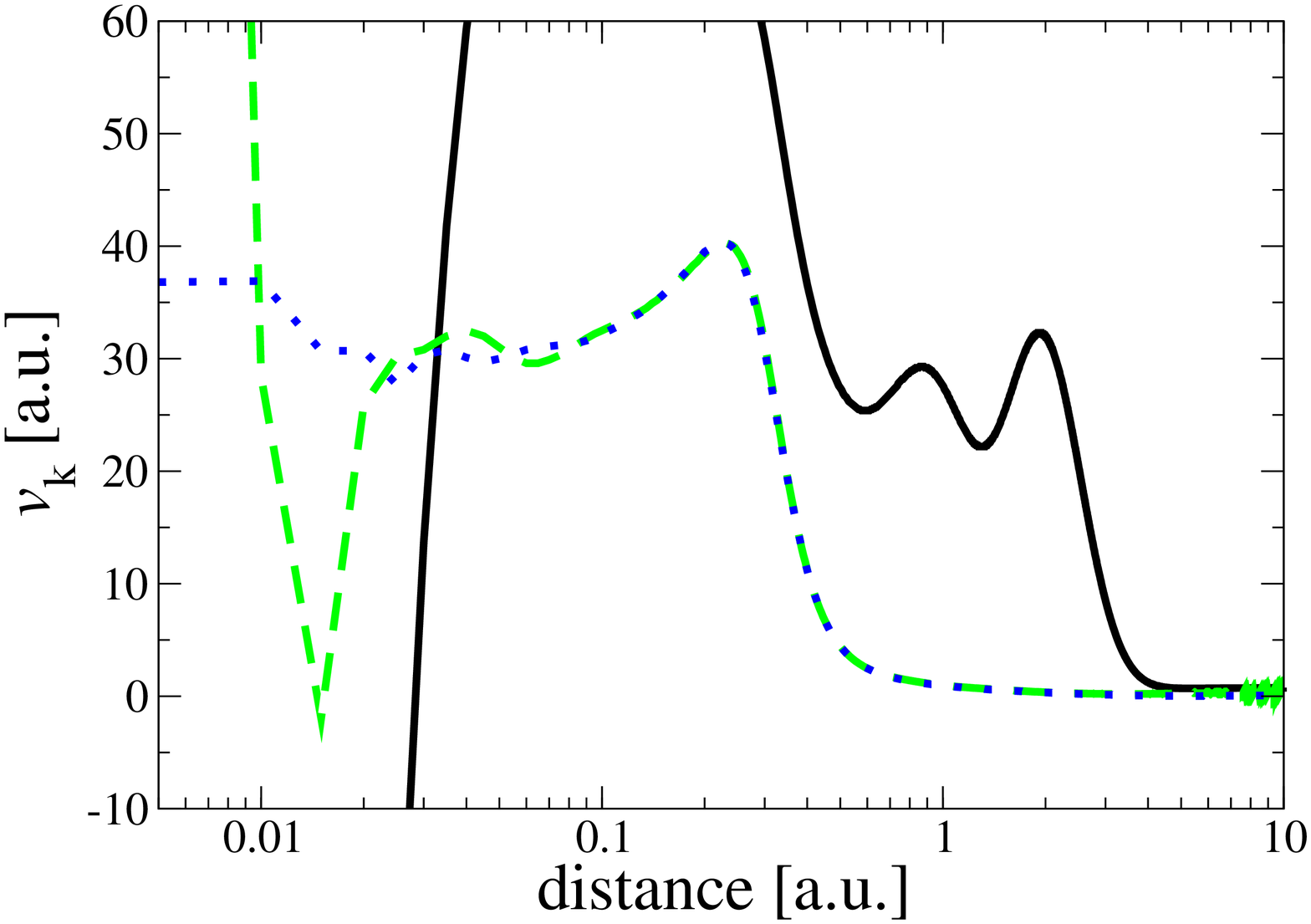}
	\caption{ Total (left) and Pauli (right) kinetic potential calculated using OEP method described in \Sec{sec:kinoep3} on top of OEPx converged quantities for Ne atom in various basis sets (see text).}
	\label{fig2}
\end{figure*}
\begin{figure}
	\includegraphics[width=1\columnwidth]{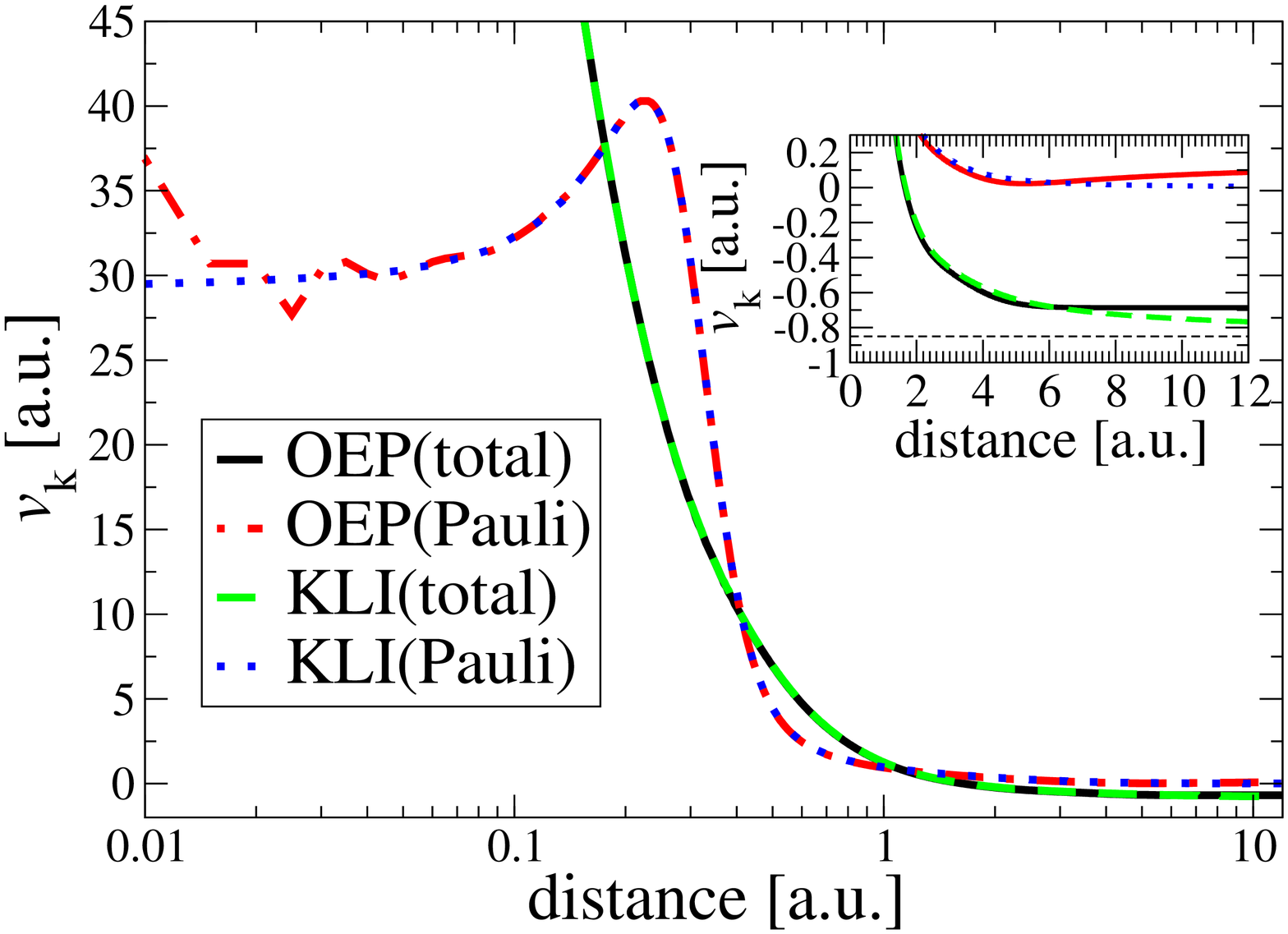}
	\includegraphics[width=1\columnwidth]{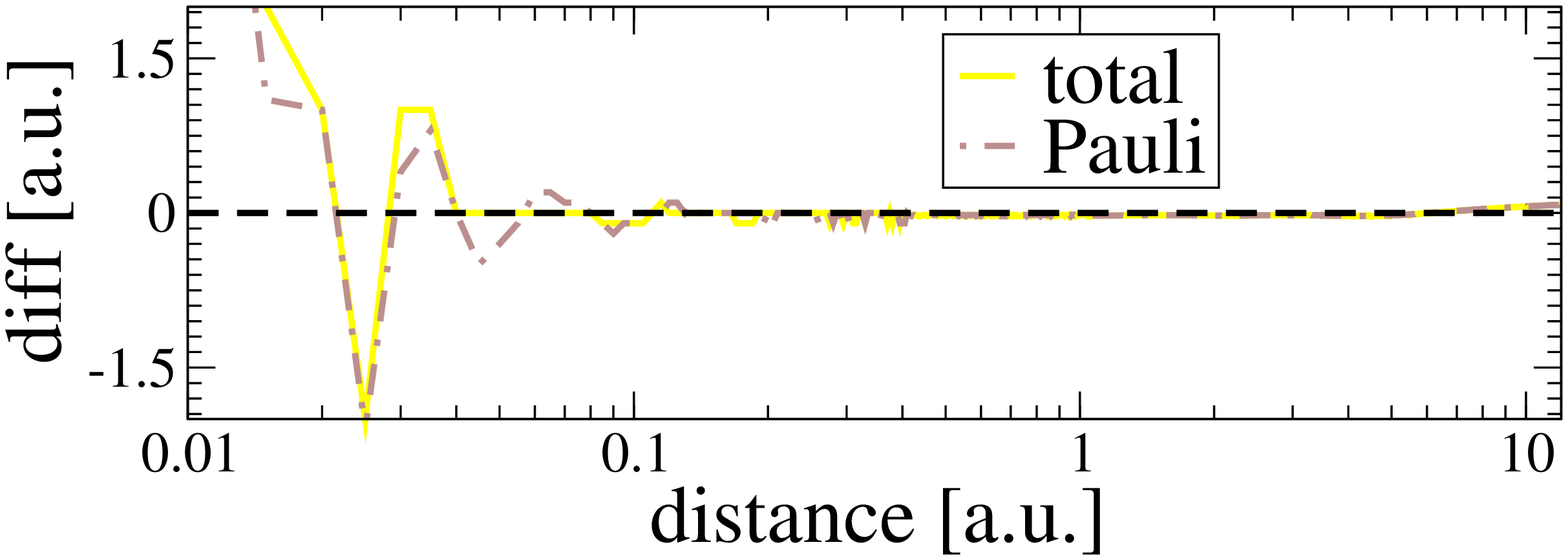}
	\includegraphics[width=1\columnwidth]{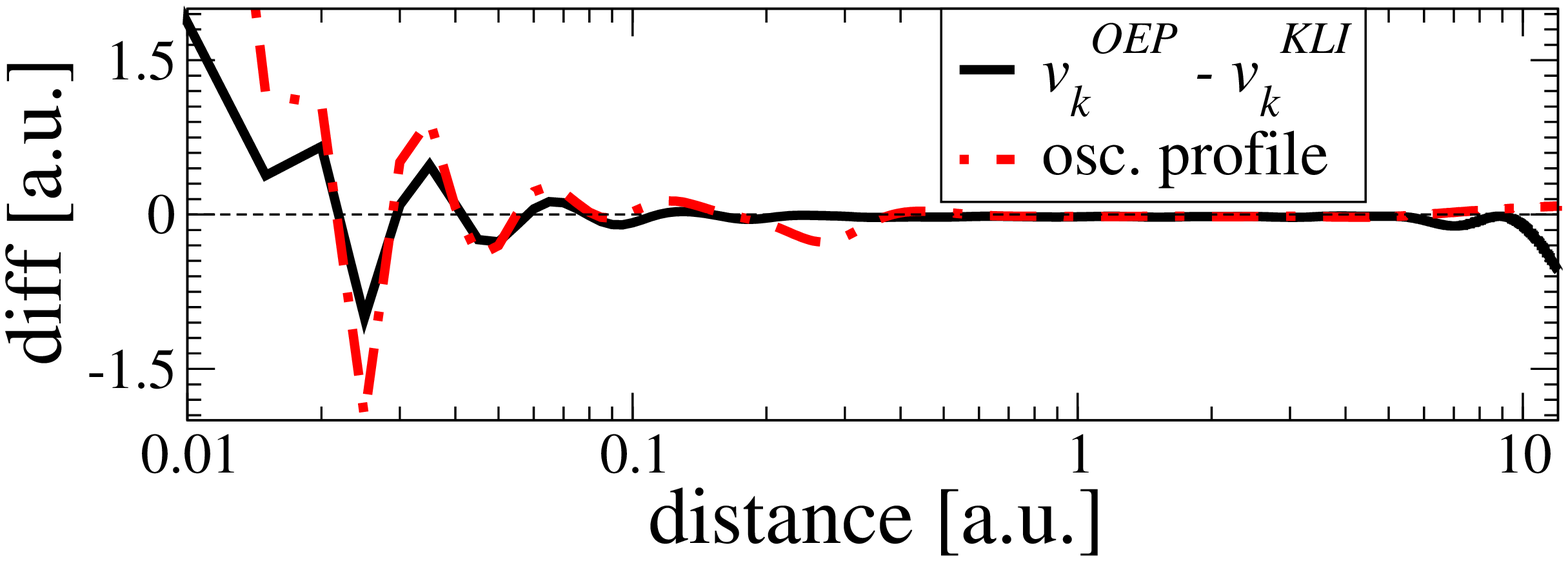}
	\caption{ (Top) comparison of total and Pauli kinetic potentials calculated using with OEP and KLI methods on top of OEPx SCF quantities using UGBS basis set. The inset presents the asymptotic behavior of potential. The dashed horizontal line denoted the HOMO energy ($\varepsilon_H = -0.8507$ a.u.).  (Middle) The difference in the total and Pauli kinetic potential between the OEP and KLI methods. (Bottom) The comparison of the difference in the total kinetic potential ($v_k^{OEPx} -v_k^{KLI}$) and the oscillatory profile (see \Eq{eq:delta1}) calculated for the same computational setup.}
	\label{fig3n}
\end{figure}
\begin{figure*}
	\includegraphics[width=1\columnwidth]{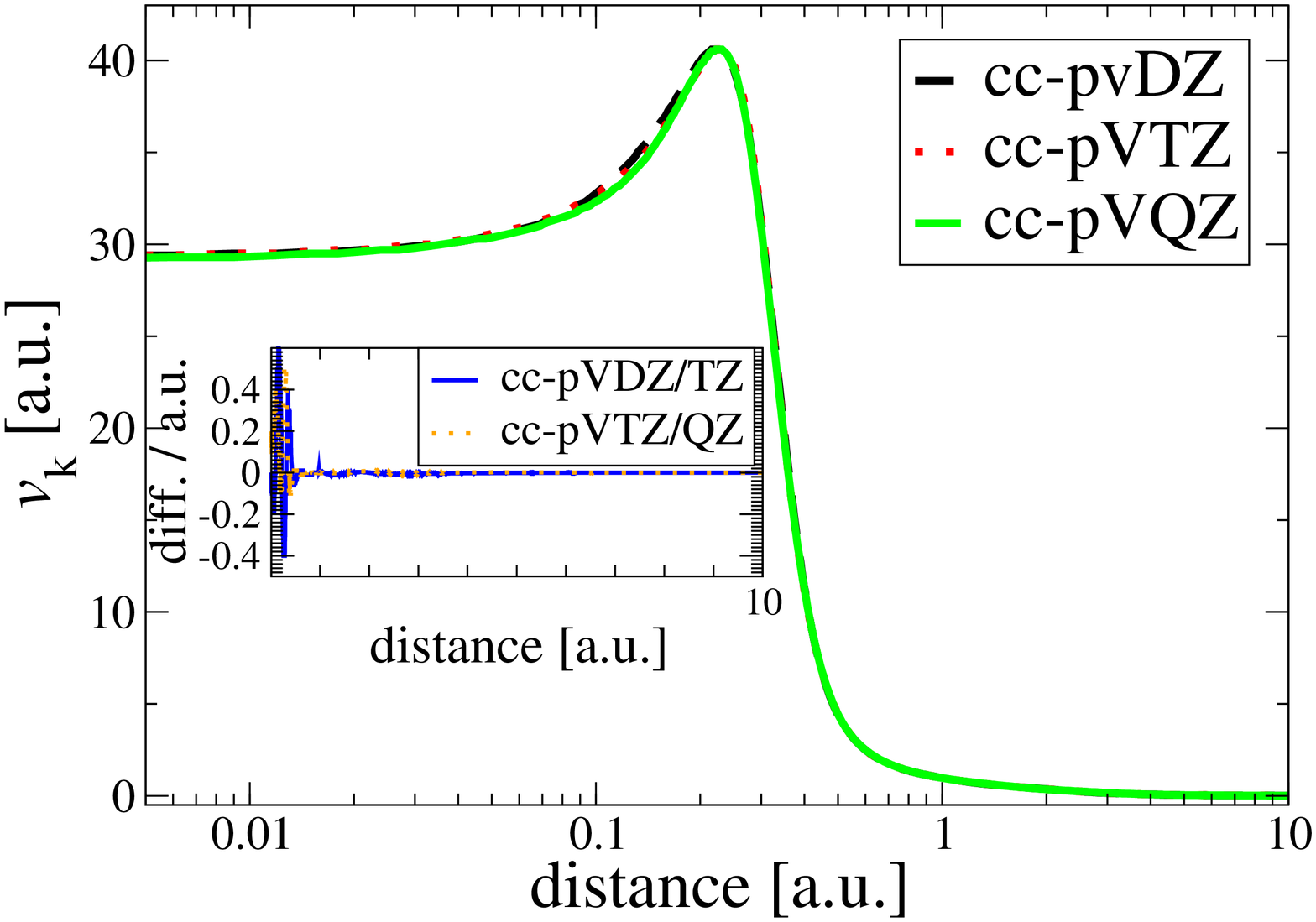}
	\includegraphics[width=1\columnwidth]{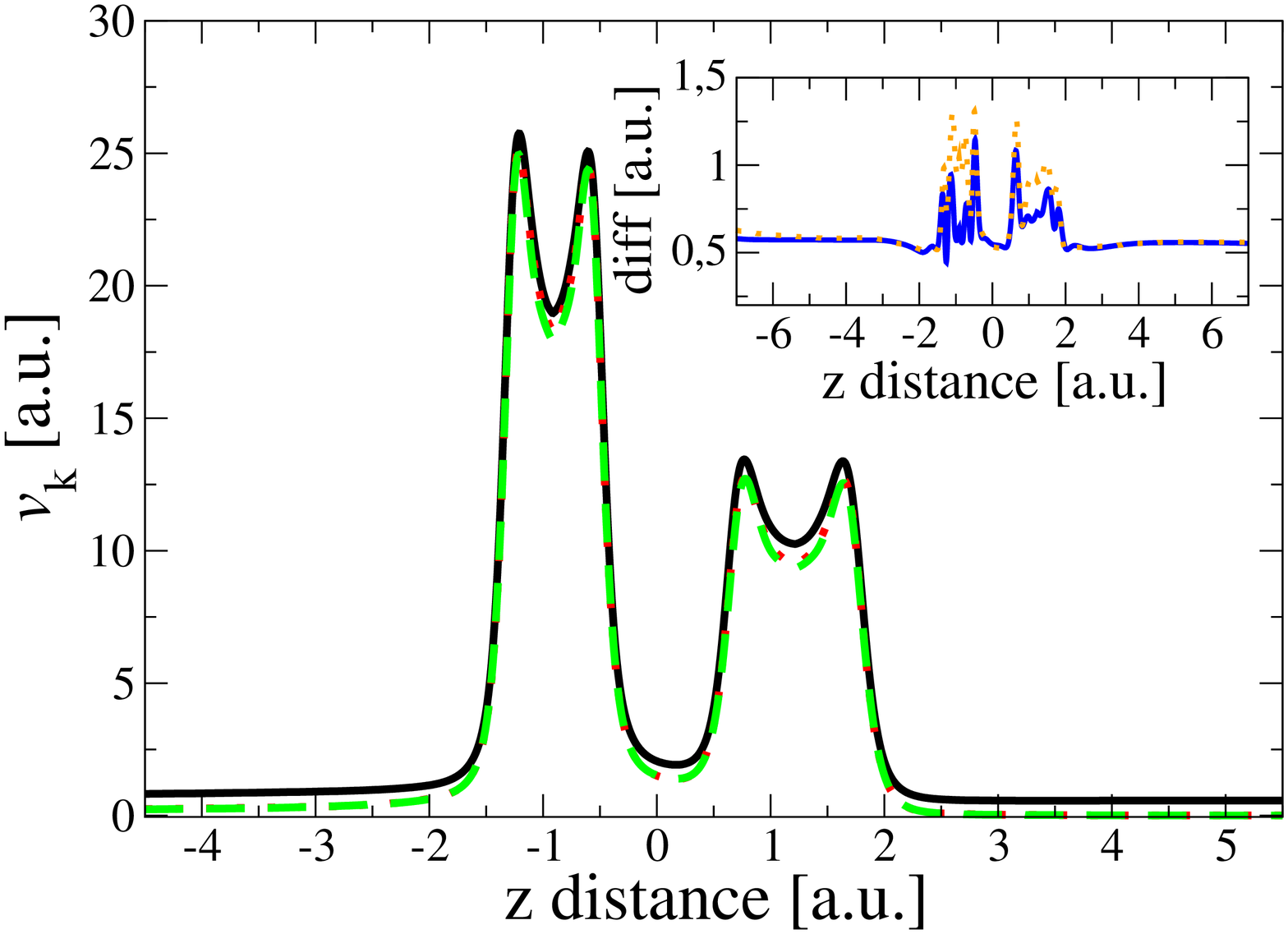}
	\caption{Pauli potential calculated using the KLI method described in \Sec{sec:kinoep4} on top of OEPx converged quantities for the Ne atom (left) and the CO molecule (right, plotted along bond axis) for the cc-pVXZ ($X = D, T, Q$) family of basis sets. The inset presents the differences between the Pauli potentials calculated for two successive basis sets in cc-pVXZ family. }
	\label{fig4n}
\end{figure*}
\begin{figure}
	\includegraphics[width=1\columnwidth]{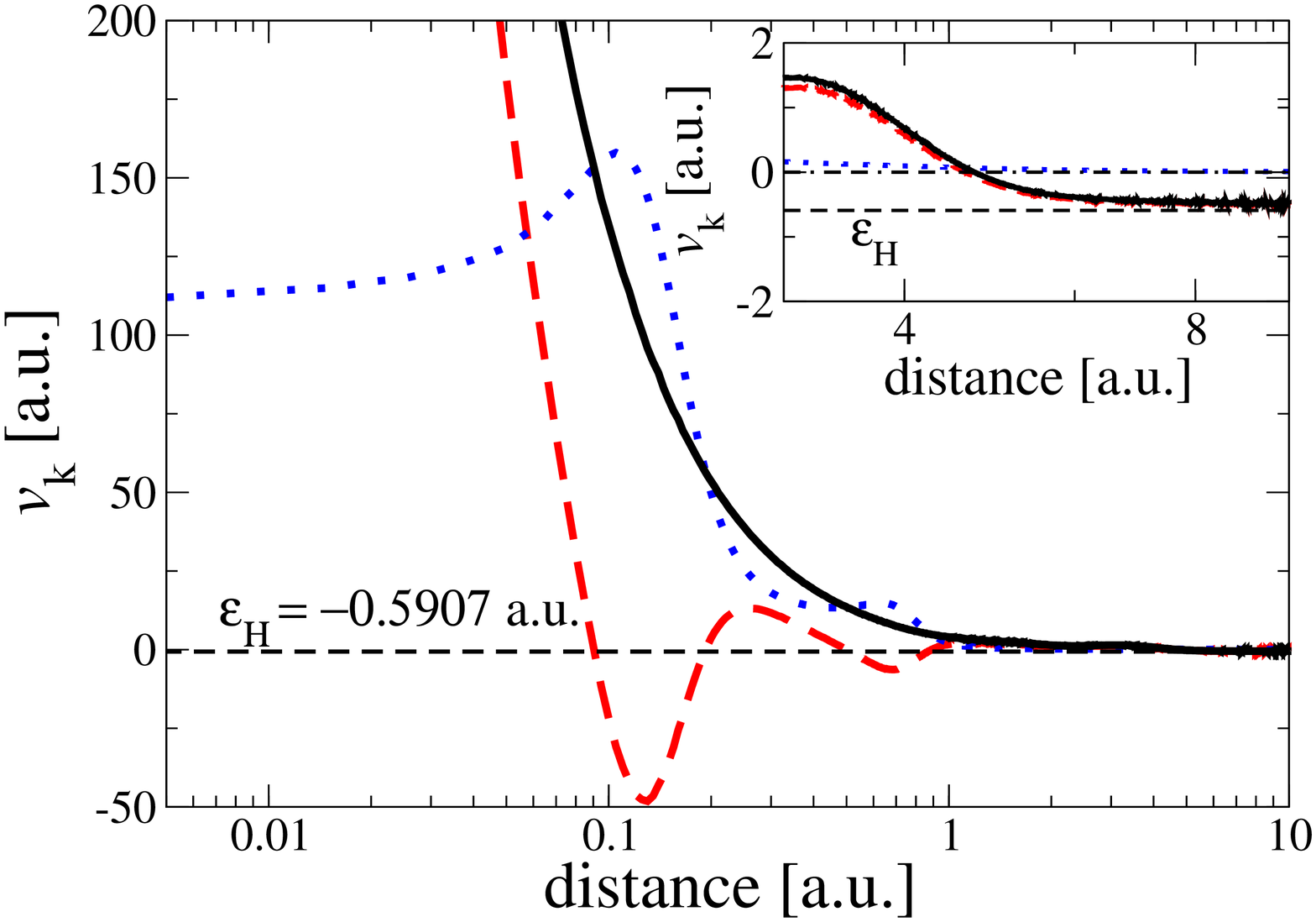}
	\includegraphics[width=1\columnwidth]{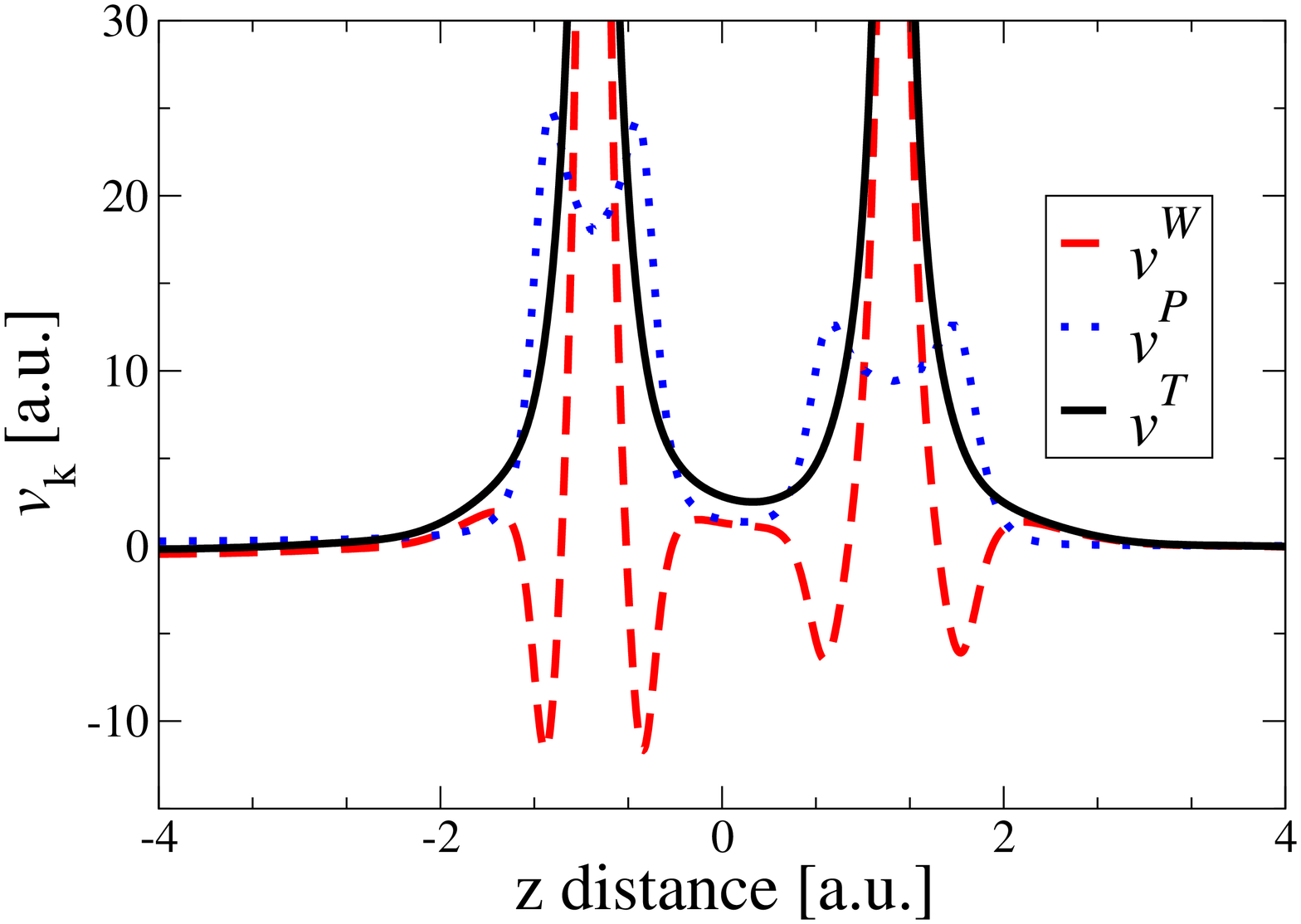}
	\includegraphics[width=1\columnwidth]{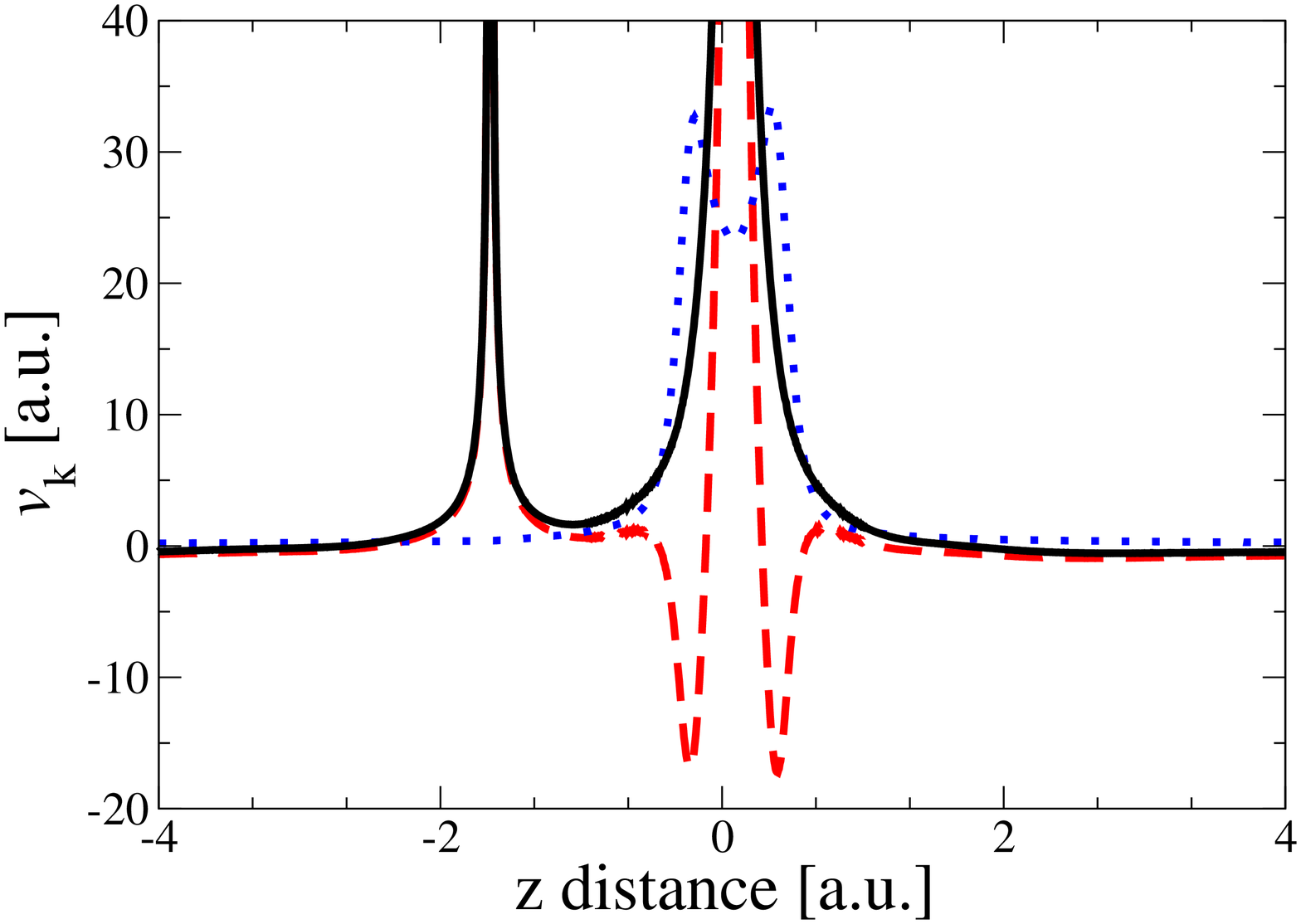}
	\caption{   Total, Weizs\"{a}cker and Pauli kinetic potentials for Ar (top) atom and CO (middle) and HF (bottom) molecules (plotted along bond axis). Data are generated on top of OEPx SCF results in cc-pVTZ basis set using the method described in \Sec{sec:kinoep4}.}
	\label{fig5n}
\end{figure}

\begin{figure*}
	\includegraphics[width=1\columnwidth]{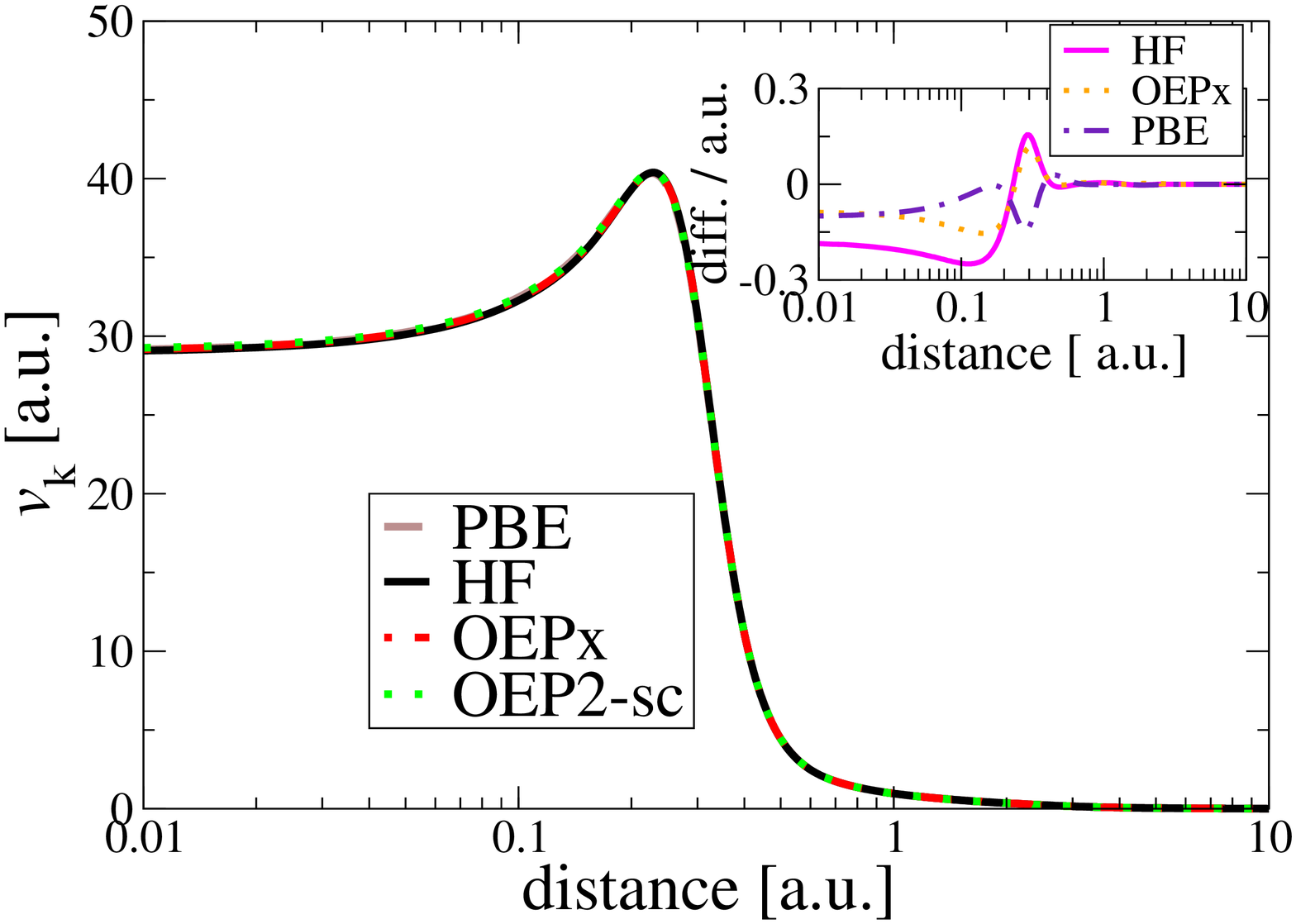}
	\includegraphics[width=1\columnwidth]{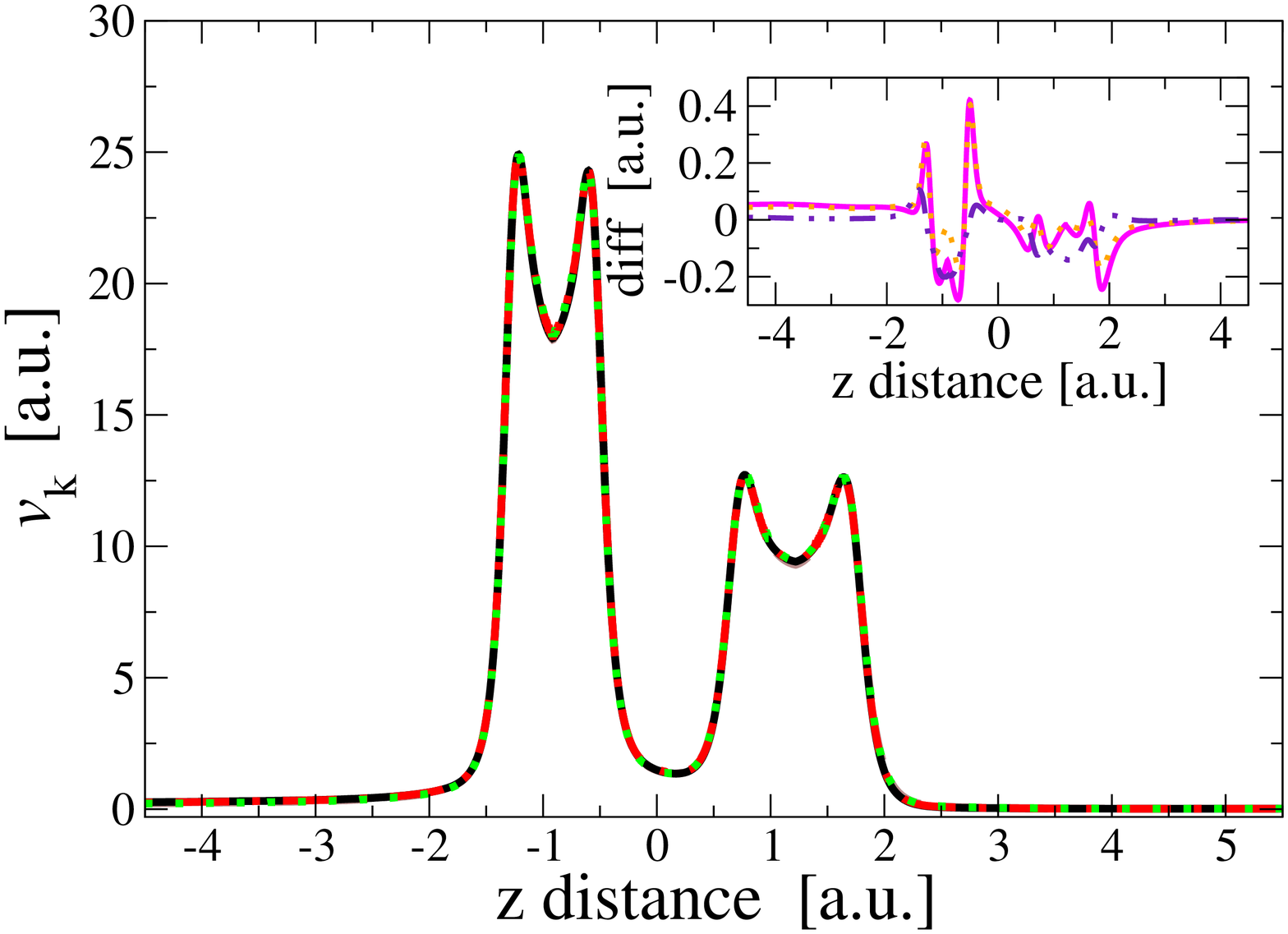}
	\caption{Pauli potential calculated using the KLI method described in \Sec{sec:kinoep4} on top of PBE, HF, OEPx and OEP2-sc converged quantities for the Ne atom (left) and the CO molecule (right, plotted along bond axis) for the cc-pVTZ basis set. The inset presents the differences between PBE, HF and OEPx results and those obtained from OEP2-sc method.}
	\label{fig6}
\end{figure*}

\section{Computational details}

 All methods have been implemented in a local version of the ACESII
 \cite{acesII} software package. The kinetic potentials have been computed for
 fixed reference densities obtained from various methods, such as
 OEPx\cite{talman:1976:OEP,ivanov:1999:OEP},
 OEP2-sc\cite{bartlett:2005:abinit2}, Hartree-Fock (HF)\cite{szabo1989modern},
 Perdew-Burke-Ernzerhof (PBE)\cite{pbe} in post-SCF fashion. 
{Here OPEx denotes exact-exchange OEP calculations, OEP2-sc denotes second-order correlated OEP calculations with a semi-canonical transformation of the orbitals.}
We remark, that a similar approach was already successfully utilized in some studies\cite{GRAB-MP-2005,Fab-JCP-2007,Fab-ISI-2019} to investigate {the} most relevant features of the XC potentials. 
To calculate the kinetic potentials with {the} OEP method, in practice, we have employed the
finite-basis set procedure of Ref. \cite{ivanov:1999:OEP}, which was also used in our previous studies to generate exchange and correlation potentials\cite{grabowski:2011:jcp,Fab-ISI-2019,DHOEPSmiga2016,GraMP2014,SCSIP,BUK-2016,Smiga2014125,LEOEP}.
Thus, the kinetic OEP potential is expanded in the same primitive Gaussian basis set which is used to represent the orbitals in the SCF procedure.
In the case of exchange and correlation potentials, this procedure led to the reduction of numerical instabilities in the solution of OEP equation\cite{hirata:2001:OEPU,hesselmanexx07,Kollmar1}. In all calculation, the 
cutoff {for the truncated singular-value decomposition} (SVD) was set to 10$^{-6}$.

In the following {subsection}, we recall some problems related to representation in real
space of von Weizs\"acker and Pauli potentials and describe the details
related to {the} implementation of the OEP {method and the KLI approximation}. 

\subsection{Asymptotic-corrected VW potential}\label{secvw}
In principle the far distance asymptotic behavior of the VW potential is
\begin{equation}
v^W (\R) \propto \varepsilon_H \; ,
\end{equation}
where $\varepsilon_H$ is {the} HOMO energy.
This property is readily obtained by using the asymptotic density behavior 
\cite{Paola-Asympt-2016} $\rho(r) \propto A e^{-2 \sqrt{-2\varepsilon_H}r}$
into Eq. (\ref{eq:vWpot}).
However, in many calculations the electron density is expanded in Gaussian 
basis functions. Thus, the asymptotic density behavior is not the true one but rather
$\rho(\R)\propto e^{-\alpha r^2}$, with $\alpha$ being the exponent of the most diffuse primitive basis function. Consequently the VW potential is found to behave as
\begin{equation}
v^W (\R) \propto -\frac{1}{2}\alpha^2r^2+\frac{1}{2}\alpha\ .
\end{equation}
Then, the VW potential incorrectly diverges. In a similar way, the use of contracted Gaussian functions to represent the electron
density may lead to oscillations in the core region of atoms, due to inaccuracies in the
description of the Laplacian term in Eq. (\ref{eq:vWpot}) \cite{Schipper1997}. These issues have been studied by several groups 
\cite{Schipper1997,Mura-JCP-1997,Jac-JCP-2011,Gaiduk-JCTC-2013,Silva-JCP-2012,Silva-PRA-2012} and are basically related to the truncation error of the basis set.

One possible remedy to reduce this basis set artifacts is to use a smoothing procedure as described in \Ref{Gaiduk-JCTC-2013}. This allows, in an effective manner, to eliminate the oscillations and divergences in the VW potential computed using 
Gaussian basis set densities by utilization of a basis-set oscillation profile 
\cite{Schipper1997,Gaiduk-JCTC-2013} defined as
\begin{equation}\label{eq:delta1}
\Delta v_{osc} (\R) = - \frac{1}{\rho(\R)} \sum^N_{i=1} \phi^*_i(\R) \delta_i(\R) 
\end{equation}
with
\begin{equation}\label{eq:delta2}
\delta_i(\R) = \left[-\frac{1}{2} \nabla^2
+v_{\text{s}}({\bf
r})[\rho] \right]\phi_{i}({\bf r}) - \varepsilon_{i}\phi_{i}({\bf r})\ .
\end{equation}
Note that, since the oscillation profile is directly linked with a measure of basis set
incompleteness \cite{Schipper1997}, for a complete basis set all $\delta_i(\R)$ vanish and thus so does the oscillation profile.

Employing the \Eq{eq:delta1} the corrected VW potential can be computed as
\begin{equation}\label{eq:vWpotfix}
v^{W,fix} (\R) = v^{W} (\R) - \Delta v_{osc} (\R) \; .
\end{equation}

\Fig{fig1} shows the VW potential of the Ne atom, computed with a Gaussian basis set (cc-pVTZ\cite{dunning:1989:bas}) with and without the oscillation profile correction. We see that the impact of the correction is observed both in core and asymptotic regions what might be very important from the computational point of view when a Gaussian-type basis set is employed in the calculation.

\subsection{Numerical implementation of OEP method}
In principle, likewise for the exchange potential, \Eq{eq:oep5} could 
be solved numerically \cite{talman:1976:OEP,EngelOEPx}. However, in
general it is better to transform the OEP equation [\Eq{eq:oep5}] into an 
algebraic problem like in \Ref{ivanov:1999:OEP,gorling:1999:OEP}.
This is done by expanding the kinetic potential and \Eq{eq:oep4a} on an auxiliary, 
orthonormal, M-dimensional basis set $\{f_p({\bf r})\}^M_{p=1}$ as 
\begin{equation}\label{eq:oep6}
v_{k}^{\mbox{\scriptsize OEP}}({\bf r})=\sum_p c_p f_p({\bf r}),
\end{equation}
and
\begin{equation}
    X({\bf r,r'})=\sum_{p,q}({\bf X})_{pq}f_p^*({\bf
    r})f_q({\bf r'}),
\end{equation}
where
\begin{eqnarray}\label{eq:oep8}
 ({\bf X})_{p q}  &&= \int f_p^*({\bf r}) X({\bf r,r'}) f_q({\bf r'}) d\R' d\R \nonumber \\
&&  =\sum_{i, a} \left (\frac{(i
a|p)(i a |q)^*}
{\varepsilon_{i} -\varepsilon_{a}}+c.c. \right ) \; ,
\end{eqnarray}
while
\begin{eqnarray}\label{eq:oep9}
\nonumber
(r s |q) = \int d{\bf r'} \phi_{s}({\bf r'}) \phi^*_{r}({\bf r'}) f_q({\bf r'}) \; .
\end{eqnarray}

This step allows to turn the solution of \Eq{eq:oep5} into 
{an} algebraic problem in which the expansion coefficients ($c_p$) 
{are} obtained from the solution of OEP equation in the form
\begin{equation} \label{oepeqmat}
     ({\bf X})_{qp}  {\bf c}_p = {\bf Y}_q \; ,
\end{equation}
with 
\begin{equation}\label{eq:oep7a}
{\bf Y}_q = \sum_{p} \sum_{i,a} \left [ \left \{ \frac{(T_s)_{i a}}
{\varepsilon_{i}-\varepsilon_{a}}(a i |q)
\right \}+c.c. \right ] \; .
\end{equation}

%
%

Note that since the density-density response matrix is singular \cite{UniqHIrata}, in order to solve \Eq{eq:oep9}  one needs to employ a truncated 
SVD
in the OEP procedure in order to calculate {the} pseudo-inverse of {the} density-density response matrix, $({\bf X}^{-1})_{qp}$, which is 
 an essential step for determining stable and physically meaningful OEP potentials \cite{UniqHIrata,ivanov:2002:OEP,grabowski:2014:jcp}.

\subsection{Numerical implementation of KLI method}
In order to obtain the total kinetic potential given by \Eq{eq:kli7} one needs to find the matrix elements ($c_i  = (v_{k})_{i i} - (T_{s})_{i i}$)  which 
{depend} explicit{ely} on {the} total kinetic potential itself. Thus, similarly as in the case of {the} KLI method applied to {the} exchange potential\cite{KLIapp,GraboOEP} one can 
{solve} this problem turning \Eq{eq:kli7} into the linear algebraic equations taking the form 
\begin{equation} \label{klieqmat}
   \left[ {\bf 1} - {\bf M} \right] {\bf c} = {\bf t}
\end{equation}
where 
\begin{equation} \label{klieqmat1}
 ({\bf M})_{kl,i} = \int d\R \frac{\phi^*_k(\R) \phi^*_i(\R) \phi_i(\R) \phi_l(\R) }{\rho(\R)}
\end{equation}
and
\begin{equation} \label{klieqmat2}
 ({\bf t})_{kl} = \bra{\phi_k}  \frac{1}{2} \nabla^2 - \frac{\tau(\R)}{\rho (\R)} + \frac{\nabla^2 \rho (\R)}{4 \rho (\R) }   \ket{\phi_l} \; .
\end{equation}

The matrix equation (\Eq{klieqmat}) can be efficiently solved using standard numerical routines  with respect to {the} $c_i$ coefficients which then can be used to compute the potential via \Eq{eq:kli7}. 

\section{Results}\label{sec:res}
In this section we show the total and Pauli kinetic potentials generated using various methods for some representative systems. More examples (for several atoms and molecules) are provided in the supporting information\cite{Supporting}. Because we have proved analytically the equivalence of the inverted Euler equation, the Bartolotti-Acharya formula and the KLI approach, in the following we will show and discuss only the OEP and KLI results.

In \Fig{fig2} we report the total (left) and Pauli (right) kinetic potentials obtained 
using the OEP procedure for the Ne atom in few basis sets.
One can note that for both kinetic potentials generated using the cc-pVTZ \cite{dunning:1989:bas} basis set, we get a nonphysical course of the potential in the asymptotic and the core region. Moreover, in the valence region, we observe strong oscillations, especially visible in the case of the Pauli potential. Similar results (not reported) were obtained also employing the cc-pVDZ and cc-pVQZ Dunning \cite{dunning:1989:bas}  basis sets. These issues are due to the fact that these basis sets are not flexible enough to represent {the} kinetic potentials and the response matrix, thus the OEP equation cannot yield a satisfactory solution. Uncontraction of the cc-pVTZ basis set (cc-pVTZU) makes it sufficiently flexible, especially in the atomic core regions, leading to a significant improvement in the shape of both potentials. In fact, in this case, the kinetic potential exhibits a much better behavior in that region.
This fact resembles what is observed in the case of the OEP procedure applied to exchange and exchange-correlation potentials\cite{UniqHIrata,hesselmanexx07,GRAB-MP-2005}. 
Nevertheless, even if the cc-pVTZU basis set definitely improves the description of the total kinetic potential, looking at the Pauli potential, which is more sensitive to numerical issues, one can still observe a moderate oscillatory behavior in the core region. This has probably the same origin as the one observed in \Ref{hesselmanexx07} in the case of exchange potentials. Those oscillations can be removed by {a} careful choice of the basis set. For example, the utilization of a larger uncontracted basis set, namely the universal Gaussian basis set \cite{UGBS} (UGBS), leads to {a} further improvement of the Pauli kinetic potential such that the oscillations are largely reduced.

\begin{table*}[htbp]
\caption{The expansion coefficients calculated using Bartolotti-Acharya formula (BA) and the one obtained from KLI method described in \Sec{sec:kinoep4} (KLI). For all systems the UGBS basis set was used.}
\begin{tabular}{cccccccc}
\hline \hline
 & \multicolumn{ 3}{c}{OEPx orbitals} & &  \multicolumn{ 3}{c}{HF orbitals} \\ 
 \cline{2-4} \cline{6-8}

orb. & orb. energy & BA ($\varepsilon_H - \varepsilon_i$)  & KLI & &  orb. energy & BA ($\varepsilon_H - \varepsilon_i$)&  KLI \\ \hline
\multicolumn{ 8}{c}{Be} \\ \hline
1s & -4.125 & 3.816 & 3.816 &  & -4.733 & 4.423 & 3.861 \\ 
2s & -0.309 & 0.000 & - &  & -0.309 & 0.000 & - \\ 
 \multicolumn{ 8}{c}{Ne} \\ \hline 
1s & -30.820 & 29.969 & 29.969 &  & -32.772 & 31.922 & 29.961 \\
2s & -1.718 & 0.867 & 0.867 &  & -1.930 & 1.080 & 0.858 \\ 
2p & -0.851 & 0.000 &  - &  & -0.850 & 0.000 & - \\ 
\hline \hline
 \end{tabular}
\label{tab:tab1}
\end{table*}

A more effective way to avoid the basis set artifacts and obtain stable and well-behaving
kinetic potentials turn{s} out to be the utilization of the KLI approximation described in \Sec{sec:kinoep4}.
In the top of \Fig{fig3n} we show in fact a comparison of the total and Pauli kinetic potentials generated using the KLI and the OEP methods (with {the} UGBS basis set). Additionally, in the middle panel of \Fig{fig3n} we report the difference in the total ($v_k^{OEPx} -v_k^{KLI}$) and Pauli kinetic potential{s} between the OEP and KLI methods. One can readily see that the KLI approximation yields virtually the same potentials as the OEP procedure but without {the} unphysical oscillations. The largest differences can be seen mostly in the core and asymptotic regions probably due to {the} 
{superposition} of two problems related to i) {the} basis set incompleteness and oscillatory profile; ii) the expansion of OEP kinetic potential in the finite Gaussian basis set. This actually can be confirm{ed} comparing the  ($v_k^{OEPx} -v_k^{KLI}$) difference with the oscillatory profile. This is reported in the bottom panel of \Fig{fig3n}. One can note that in the major part these two quantities are largely proportional to each other meaning that the difference between the kinetic potential generated by {the} OEP and {the} KLI methods lays basically in the incompleteness of the basis set used to expand both orbitals (thus {the} density) and the kinetic potential. Moreover, the KLI approximation appears to incorporate
the oscillation profile correction thus the quality of the total and Pauli potentials are much better. For example, in the asymptotic region, the KLI method performs much better than the OEP potential (see the inset in the upper panel) which is not decaying correctly to $\varepsilon_H$ for large values or $r$ (and as $1/r^2$ \cite{Const-PRB-2019} to zero in case of {the} Pauli potential). This, in fact, is related to the Gaussian basis (used to expand the OEP kinetic potential) which goes rapidly to zero in this region. {A}
{s}imilar behavior was also observed in the case of{the}  exchange OEP potential\cite{hirata:2001:OEPU,UniqHIrata,ivanov:2002:OEP}. Moreover, we note that, in the case of {the} KLI approximation, the description of the core region is highly improved. The occurrence of rapid oscillation{s} in the OEP potential in this region has probably the same origin as the one observed in \Ref{hesselmanexx07} in the case of {the} exchange potential
and can be cured by {a} proper balancing the auxiliary basis set in OEP procedure. 

One more advantage of the KLI method is that the quality of the results is preserved also 
when standard, relatively small basis sets are used. This is shown in 
\Fig{fig4n} where we report the Pauli potentials for Ne atom and CO molecule generated using the family of Dunning~\cite{dunning:1989:bas} cc-pV$X$Z basis sets (where $X = D, T, Q$). 
The plots show that indeed the potential is only marginally dependent on the basis set and
in any case no numerical artifacts appear. This shows that the KLI approach for the description of the kinetic potential is really a robust numerical procedure.

In \Fig{fig5n} we report the total, von Weizs\"{a}cker, and Pauli kinetic potentials generated using the KLI method and the cc-pVTZ~\cite{dunning:1989:bas} basis set for three representative systems, namely the Ar (top) atom and the CO (middle) and HF (bottom) molecules. The same quantities are reported for several other atomic and molecular systems in the supporting materials (see \Ref{Supporting}). First of all, we note the smooth course of all kinetic potentials. This further supports the conclusion that the KLI method is stable and can generate reference potentials for any type of system. Secondly, the Pauli potential is always non-negative, giving a finite value at the core\cite{PauliCusp} and decays to zero asymptotically\cite{Const-PRB-2019}. Moreover, at the bond (see e.g. the HF and CO cases, as well as other molecules in \Ref{Supporting}) the Pauli potential gives a non-negative contribution to the total kinetic potential which may play quite an important role in some cases. Furthermore, we see that in the iso-orbital regions {the} Pauli potential goes correctly to zero (see e.g. the H side in the HF molecule). The Pauli potential also exhibits {a} similar shell structure as the one visible in the Weizs\"{a}cker kinetic potential case. We note that the total kinetic potential does not have such features meaning that those must almost cancel mutually. This is {an} important fact which should be taken into account in the construction of KE functionals and potentials for OF-DFT. Finally, we note that {the} total and Weizs\"{a}cker kinetic potentials show {the} correct behavior at the nuclei and in the asymptotic region where they decay to $\varepsilon_H$. (see the insets in \Fig{fig5n}, Ar atom).

Finally, we have assessed the impact of {the} reference SCF orbitals on the shape of total and Pauli kinetic potentials. Thus in \Fig{fig6} we report the aforementioned quantities calculated on top of OEPx\cite{talman:1976:OEP,ivanov:1999:OEP}, OEP2-sc\cite{bartlett:2005:abinit2}, HF\cite{szabo1989modern} and PBE\cite{pbe} orbitals in the post-SCF fashion for Ne atom and CO molecule in the cc-pVTZ basis sets~\cite{dunning:1989:bas}. 
One can readily see that all {the} orbitals generate virtually the same Pauli potentials. 
{A} closer look (see the inset of \Fig{fig6} where we present the differences between PBE, HF and OEPx results and the one obtained from OEP2-sc method) reveals that some differences appear mostly in the core regions. Note, that the Pauli potential obtained from HF orbitals also behaves in line with others. This is somehow contradictory to the results reported in \Ref{Finzel2018} (see Fig. 2 in the paper). However, the closer inspection reveals that the Pauli potential in \Ref{Finzel2018} was obtained using {the} BA formula which, in contrary to our KLI method, incorporates also {the} effects related to the non-locality of exchange operator (see \App{ap:ABF3} for more details). This can be also seen in \Tab{tab:tab1} where we report the expansion coefficients calculated using {the} BA formula ($\varepsilon_H - \varepsilon_i$) and {the} KLI method for Be and Ne atom in UGBS basis set. One can immediately note that in the case of {the} OEPx orbitals the expansion coefficients are identical. On the other hand, in the case of {the} HF orbitals, there is a quite large discrepancy between {the} coefficients. As was mentioned before, this is due to the fact that the BA coefficients obtained for {the} HF orbitals take also into account the energy shift ($\bra{\phi^{HF}_i} v_{x} -  \hat{v}^\text{NL}_{x} \ket{\phi^{HF}_i}$) related to the difference between {the} local and non-local exchange potential{s}. In {the} KLI case the utilization of HF quantities lead{s} to a purely local kinetic potential which does not include any additional effect
related to {the} exchange potential (see \App{ap:ABF3}) and thus the coefficients are much more similar to the one{s} obtained from the OEPx method.

\section{Conclusions}

{We have introduced a new method that allows to generate a real-space representation of the total and Pauli kinetic potentials
via {the} utilization of {the} OEP method \cite{sharp:1953:OEP,talman:1976:OEP} taking as
a starting point the KS non-interacting kinetic energy expression. Moreover,
we have reviewed in detail all {the} presently utilized methods used to compute {the}
aforementioned quantities. The OEP based method, however,} leads to similar
numerical problems 
{as those} encountered in the case of {the} exchange and
exchange-correlation OEP calculations. Thus, we have derived a common energy
denominator approximation to the kinetic OEP method 
{and then its KLI variant}
which has proven to give much more stable and robust results than the original OEP one.
Additionally, we have proved that 
{when the SCF density and orbitals are employed,}
at the solution point{,} our KLI method is formally equivalent to commonly used BA formula \cite{Bar-JCP-1982} when KS reference orbitals are considered. Nevertheless, our KLI approach seems to be superior to the BA one because it can be also employed starting from HF orbitals (whereas in this case, the BA formula includes some undesired non-local contributions).  We hope that the present work will shed some light on the future development of total and Pauli kinetic potentials for OF-DFT and allow  to find some new or improve existing \cite{constantin2018semilocal} semilocal expressions for the latter.

%
%

\section*{Acknowledgments}
This work was supported by the Polish National Science Center under Grant No. 2016/21/D/ST4/00903. 

\appendix

\section{Bartolotti-Acharya formula}\label{ap:ABF} 
Consider the KS equation
\begin{equation}
\left[-\frac{1}{2}\nabla^2+v_{s}(\R)\right]\phi_{i}(\R)=\varepsilon_{i}\phi_{i}(\R)\ .
\end{equation}
Multiplication by $\phi^*_i$ and sum over $i$ yields
\begin{equation}\label{eq:D1}
\frac{\tau_L({\bf r})}{\rho({\bf r})}  +v_{\text{s}}(\R) =  \frac{1}{\rho({\bf r})} \sum^N_i  \varepsilon_{i} |\phi_{i}({\bf r}) |^2 \ ,
\end{equation}
where $\tau_L(\R)=-(1/2)\sum_i\phi_{i}^*(\R)\nabla^2\phi_{i}(\R)$.
Using Eq. (\ref{tauleq}) and the fact that $\tau(\R)=\tau^W(\R)+\tau^P(\R)$, we find

\begin{equation}\label{eq:D2}
\frac{\tau^W}{\rho} - \frac{\nabla^2 \rho}{4 \rho} + \frac{\tau^P}{\rho} + v_{\text{s}}(\R) =  \frac{1}{\rho({\bf r})} \sum^N_i  \varepsilon_{i} |\phi_{i}({\bf r}) |^2 \ .
\end{equation}

On the other hand, the Euler equation [Eq. (\ref{eq:euler})] reads
\begin{equation}\label{eq:D3}
\frac{\delta T_s[\rho]}{\delta \rho(\R)} + v_{\text{s}}(\R) = \mu \; .
\end{equation}
By subtracting \Eq{eq:D2} from \Eq{eq:D3}, after some algebra we obtain
\begin{equation}\label{eq:D4}
\frac{\delta T_s[\rho]}{\delta \rho(\R)} =   \frac{\tau^W}{\rho} - \frac{\nabla^2 \rho}{4 \rho} + \frac{\tau^P}{\rho}  + \frac{1}{\rho({\bf r})} \sum^N_i \left( \mu - \varepsilon_{i} \right)|\phi_{i}({\bf r}) |^2   \; ,
\end{equation}
Now if we remove the von Weizs\"acker kinetic potential 
from both sides we easily retrieve \Eq{eq:BA1}.

It is also easy to show that the Bartolotti-Acharya formula can be
alternatively derived starting from the oscillation profile corrected Pauli potential
\begin{equation}\label{eq:vkin1pfix}
v^{P,fix} (\R) = -v_s(\R) - v^{W} (\R) + \Delta v_{osc} (\R) + \mu\ ,
\end{equation}
that is directly derivable from Eq. (\ref{eq:vWpotfix}). 
Taking into account that the $\Delta v_{osc} (\R) $ can be rewritten as
\begin{eqnarray}\label{eq: D2}
\nonumber
\Delta v_{osc} (\R) && =  \frac{\tau^W(\R)}{\rho(\R)} - \frac{\nabla^2 \rho(\R)}{4 \rho(\R)} + \frac{\tau^P(\R)}{\rho(\R)}  + v_{\text{s}}(\R)  \\ && - \frac{1}{\rho({\bf r})} \sum^N_i \varepsilon_{i} |\phi_{i}({\bf r}) |^2\; 
\end{eqnarray}
one arrives at 
\begin{equation}\label{eq: D3}
v^{P} (\R)  =  \frac{\tau^P(\R)}{\rho(\R)}    - \frac{1}{\rho({\bf r})} \sum^N_i \varepsilon_{i} |\phi_{i}({\bf r}) |^2 +  \mu
\end{equation}
that is exactly \Eq{eq:BA1}.

\section{Equivalence of Bartolotti-Acharya formula and KLI approximation}\label{ap:ABF3} 
From the KS equation we easily find 
\begin{equation}\label{eq:E1}
\varepsilon_i = \bra{\phi_i}  -\frac{1}{2} \nabla^2 + v_{\text{s}} \ket{\phi_i} \; .
\end{equation}
On the other hand, multiplying the Euler equation [\Eq{eq:D3}] 
by $|\phi_i(\R)|^2$ and integrating over {the} whole space one arrives at
\begin{equation}\label{eq:E2}
\mu = \bra{\phi_i}  \frac{\delta T_s[\rho]}{\delta \rho(\R)} + v_{\text{s}} \ket{\phi_i} \; .
\end{equation}
Now, subtracting \Eq{eq:E1} and  \Eq{eq:E2} we get
\begin{equation}\label{eq:E3}
\mu - \varepsilon_i = \bra{\phi_i}  \frac{\delta T_s[\rho]}{\delta \rho(\R)}  + \frac{1}{2} \nabla^2 \ket{\phi_i}  = (v_{k})_{ii} - (T_{s})_{ii}\; .
\end{equation}
Finally, inserting \Eq{eq:E3} into \Eq{eq:D4} and removing from both sides the VW potential, we recover \Eq{eq:kli8}. 

Alternatively starting from {the} HF equations we find
\begin{equation}\label{eq:E4}
\varepsilon^{HF}_i = \bra{\phi^{HF}_i}  -\frac{1}{2} \nabla^2 + v_{\text{ext}} + v_\text{H} + \hat{v}^\text{NL}_{x} \ket{\phi^{HF}_i} \; ,
\end{equation}
where $\hat{v}^\text{NL}_{x}$ is a non-local HF exchange operator.
Multiplying now the Euler equation [\Eq{eq:D3}] 
by $|\phi^{HF}_i(\R)|^2$ and integrating over {the} whole space one arrives at
\begin{equation}\label{eq:E5}
\mu = \bra{\phi^{HF}_i}  \frac{\delta T_s[\rho]}{\delta \rho(\R)} + v_{\text{s}} \ket{\phi^{HF}_i} \; .
\end{equation}
Subtracting \Eq{eq:E4} and \Eq{eq:E5} we get
\begin{eqnarray}\label{eq:E6} 
\nonumber
\mu - \varepsilon^{HF}_i && = \bra{\phi^{HF}_i}  \frac{\delta T_s[\rho]}{\delta \rho(\R)}  + \frac{1}{2} \nabla^2 \ket{\phi^{HF}_i} \\ && + \bra{\phi^{HF}_i} v_{x} -  \hat{v}^\text{NL}_{x} \ket{\phi^{HF}_i} \; . 
%
\end{eqnarray}
Finally, inserting \Eq{eq:E6} into \Eq{eq:D4} and removing from both sides the VW potential one obtain{s} the expression for {the} Pauli kinetic  potential. Note, however, that \Eq{eq:kli8} does not include {the} $\bra{\phi^{HF}_i} v_{x} -  \hat{v}^\text{NL}_{x} \ket{\phi^{HF}_i}$ term which additionally accounts in {the} $\mu - \varepsilon^{HF}_i$ difference for the non-local effect related to {the} HF exchange operator.

\section{Derivation of $\frac{\delta T_{s}[\{\phi_{q}\}]} {\delta \phi_{p}({\bf r'})} $}\label{ap:GLK} 
The derivative of the orbital-dependent KE functional given by \Eq{eq:oep3} with respect to {the} orbitals reads
 \begin{widetext}
\begin{eqnarray}\label{eq:F1}\
\frac{\delta T_{s}[\{\phi_{q}\}]} {\delta \phi_{p}({\bf r'})}  & =& -\frac{1}{2}  \sum^N_i \int d\R  \frac{\delta \phi^*_{i}(\R) } {\delta \phi_{p}({\bf r'})}  \ \nabla_r^2  \phi_{i} (\R) - \frac{1}{2}   \sum^N_i \int d\R  \phi^*_{i}(\R)  \ \nabla_r^2  \left( \frac{\delta \phi_{i}(\R) } {\delta \phi_{p}({\bf r'})} \right) \nonumber \\
& =& -\frac{1}{2}   \sum^N_i \int d\R  \delta(\R-\R')\delta_{pi}  \nabla_r^2  \phi_{i} (\R)  -\frac{1}{2}  \sum^N_i \int d\R  \delta_{pi} \phi^*_{i}(\R)   \nabla_r^2  \Big( \delta(\R-\R') \Big)\ .
\end{eqnarray}
 \end{widetext}
{Clearly, whenever $\phi_p$ denotes an unoccupied orbital, we have $\delta
  Ts/\delta\phi_p = 0$. In the opposite case ($\phi_p$ being an occupied
  orbital), a non-zero value is obtained.}
After some algebra, \Eq{eq:F1} reduces to 

\begin{eqnarray}\label{eq:F2}\
\frac{\delta T_{s}[\{\phi_{q}\}]} {\delta \phi_{p}({\bf r'})} 
& =& -\frac{1}{2}   \ \nabla_r^2  \phi_{p} (\R') -\frac{1}{2}  \int d\R   \phi^*_{p}(\R)  \ \nabla_r^2  \Big( \delta(\R-\R') \Big)  .
\end{eqnarray}

Now, utilizing {the} following relation 
\begin{equation}\label{eq:F3}
\int d\R' f(\R')  {\nabla^2}_{r'} \delta(\R-\R') =  \int d\R' \delta(\R-\R') \nabla^2_{r'}  f(\R') = \nabla^2_r f(\R) \; ,
\end{equation}
on the right hand side of \Eq{eq:F2} we recover \Eq{eq:oep3}.

\bibliography{kin-oep}
\end{document}